\documentclass[12pt]{article}
\usepackage{amssymb,latexsym}
\def\ZZ{{\mathbb Z}}
\def\QQ{{\mathbb Q}}
\def\RR{{\mathbb R}}
\def\CC{{\mathbb C}}

\newtheorem{formula}{}[section]

\newtheorem{definition}[formula]{\indent Definition}
\newtheorem{corollary}[formula]{\indent Corollary}
\newtheorem{remark}[formula]{\indent Remark}
\newtheorem{lemma}[formula]{\indent Lemma}
\newtheorem{theorem}[formula]{\indent Theorem}

\def\thrm{\begin{theorem}}
\def\thrml#1{\begin{theorem}\label{#1}}
\def\ethrm{\end{theorem}}
\def\rmrk{\begin{remark}}
\def\rmrkl#1{\begin{remark}\label{#1}}
\def\ermrk{\end{remark}}
\def\dfntn{\begin{definition}}
\def\dfntnl#1{\begin{definition}\label{#1}}
\def\edfntn{\end{definition}}
\def\nmrt{\begin{enumerate}}
\def\enmrt{\end{enumerate}}

\def\qtn{\begin{equation}}
\def\qtnl#1{\begin{equation}\label{#1}}
\def\eqtn{\end{equation}}
\def\lmm{\begin{lemma}}
\def\lmml#1{\begin{lemma}\label{#1}}
\def\elmm{\end{lemma}}
\def\crllr{\begin{corollary}}
\def\crllrl#1{\begin{corollary}\label{#1}}
\def\ecrllr{\end{corollary}}

\begin{document}
\title{}
\date{}
\maketitle
%\nopagenumbers
%\begin{titlepage}
%\title
\vspace{-0,1cm} \centerline{\bf POLYNOMIAL COMPLEXITY RECOGNIZING A
TROPICAL LINEAR VARIETY}
%\centerline{\bf AND RELATIVE KOLMOGOROV COMPLEXITY}
\vspace{7mm}
\author{
\centerline{Dima Grigoriev}
%\\[-1pt]
\vspace{3mm}
%\centerline{$^1$ St.Petersbourg University,
% Universitetskaya nab., 7/9,}
%\centerline{ St.Petersbourg,
% 199164, RUSSIA}
%\vspace{3mm}
%Dima Grigoriev \\[-1pt]
\centerline{CNRS, Math\'ematique, Universit\'e de Lille, Villeneuve
d'Ascq, 59655, France} \vspace{1mm} \centerline{e-mail:\
dmitry.grigoryev@math.univ-lille1.fr } \vspace{1mm}
\centerline{URL:\ http://en.wikipedia.org/wiki/Dima\_Grigoriev} }
%\date{}
%\maketitle

\begin{abstract}
A polynomial complexity algorithm is designed which tests whether a
point belongs to a given tropical linear variety.
\end{abstract}

\section*{Introduction}

Consider a linear system \begin{eqnarray} \label{1} A\cdot X=b
\end{eqnarray} with the $m\times n$ matrix $A=(a_{i,j})$ and the vector
$b=(b_i)$  defined over the field
$K=\CC((t^{1/\infty}))=\{c=c_0t^{i_0/q}+c_1t^{(i_0+1)/q}+\cdots\}$
of Puiseux series where $i_0\in \ZZ,\, 1\le q\in \ZZ$. Consider the
map $Trop(c)=i_0/q\in \QQ$ and $Trop(0)=\infty$. Denote by $P\subset
K^n$ the linear plane determined by the system (\ref{1}). The
closure in the euclidean topology $\overline{Trop (P)} \subset
\RR^n$ is called a {\it tropical linear variety} \cite{SS} (for the
basic concepts of the tropical geometry see \cite{I}, \cite{M}).

More generally, the {\it tropical variety} attached to an ideal
$I\subset K[X_1,\dots,X_n]$ is defined as $\overline{Trop(U)}
\subset \RR^n$ where $U\subset K^n$ is the variety determined by
$I$. A {\it tropical basis} of $I$ is a finite set $f_1,\dots,f_k\in
I$ such that $\overline{Trop(U)}=\overline{Trop(V(f_1))} \cap \cdots
\cap \overline{Trop(V(f_k))}$ where $V(f_1)\subset K^n$ denotes the
variety of all the zeroes of $f_1$. In \cite{S} an algorithm is
devised which produces a tropical basis of an ideal. Having a
tropical basis, one can easily test, whether a point $v\in \RR^n$
belongs to the tropical variety $\overline{Trop(U)}$ since
$\overline{Trop(V(f ))}=V(Trop(f))$ (due to \cite{K}) where the
tropicalization $Trop(f)$ of a polynomial $f\in K[X_1,\dots,X_n]$ is
defined coefficientwise, and $V(Trop(f))$ is the tropical
hypersurface of all the zeroes of the tropical polynomial $Trop(f)$.
But on the other hand, in \cite{S} an example is exhibited of a
tropical linear variety with any its tropical basis having at least
an exponential number of elements, while the algorithm recognizing a
tropical linear variety designed in the present paper has the
polynomial complexity.

We study the problem of the complexity of recognizing
$\overline{Trop (P)}$. In other words, we design a polynomial
complexity algorithm which for a given vector $v=(v_1,\dots,v_n)\in
(\RR \cap \overline{\QQ})^n$ with real algebraic coordinates tests,
whether a system (\ref{1}) has a solution $x=(x_1,\dots,x_n) \in
K^n$ with $Trop(x)=v$. Obviously, this captures also the case $v\in
((\RR \cap \overline{\QQ})\cup \{\infty\})^n$, so we can w.l.o.g.
assume below that $v\in (\RR \cap \overline{\QQ})^n$.

Observe that the problem of recognizing, whether just the zero
vector belongs to a given tropical (non-linear) variety, is
$NP$-hard (see \cite{J}) since the solvability of a system of
polynomial equations from $\overline{\QQ} [X_1,\dots,X_n]$ is
equivalent to that the zero vector belongs to the tropical variety
determined by this system.

We mention also that testing emptiness of a tropical {\it non-linear
prevariety} (i.~e. an intersection of a few tropical hypersurfaces)
is $NP$-complete \cite{T}, while testing the emptiness of a tropical
{\it linear prevariety} belongs to $NP \cap coNP$ (\cite{A},
\cite{B}, \cite{G}).

\section{Algorithm lifting a point to a tropical linear variety}

We assume that the entries $a_{i,j},b_i\in K,\, 1\le i\le m,\, 1\le
j\le n$ are provided in the following way (cf. \cite{C}, \cite{G83},
\cite{R}). A {\it primitive element} $z\in K$ is given as a root of
a polynomial equation $h(t,z)=0$ where $h\in \ZZ[t,Z]$, and by means
of a further specifying a beginning  of the expansion of $z$ as a
Puiseux series over the field $\overline{\QQ}$ of algebraic numbers
(to make a root of $h$ to be unique with this beginning of the
expansion). Also we are supplied with rational functions $h_{i,j},
h_i \in \QQ (t)[Z]$ such that $a_{i,j}=h_{i,j}(z),\, b_i=h_i(z),\,
1\le i\le m,\, 1\le j\le n$. We suppose that $\deg(h),
\deg(h_{i,j}), \deg(h_i) \le d$. In addition, we assume that each
rational coefficient of the polynomials $h,h_{i,j},h_i$ is given as
a quotient of a pair of integers with absolute values less than
$2^M$. The latter means that the bit-size of this rational number is
bounded by $2M$.

First, the algorithm cleans the denominator in the exponents of the
Puiseux series of $z$ replacing $t^{1/q}$ by $t$ for a suitable
$q\le d$ to make $z$ to be a Laurent series with integer exponents
(and keeping the same notation for $z,\, h,\, h_{i,j},\, h_i$). Here
and below to develop a Puiseux series with  coefficients from
$\overline{\QQ}$ within the polynomial (in the bit-size of the input
and in the number of terms of the expansion) complexity, we exploit
the algorithm from \cite{C86}. The algorithm makes use of a
presentation of the field of the coefficients as a finite extension
of $\QQ$ via its primitive element (see e.~g. \cite{C}, \cite{G83},
\cite{R}) similar to the presentation of $a_{i,j}, b_i$ above (we
don't dwell here on the details since it does not influence  the
main body of the algorithm, for the sake of simplifying the
exposition a reader can suppose that the coefficients of the Puiseux
series of $z$ are rational).

The coordinates of the vector $v$ we also multiply by $s$ and keep
the same notation for $v$. We say that two coordinates $v_{j_1},\,
v_{j_2}$ of $v$ are {\it congruent} if $v_{j_1}-v_{j_2}\in \ZZ$.
Every class of congruence provides a subsystem of (\ref{1}) (note
that just one of such classes can contain the column $b$, namely the
class of $v_j$ being integers). Observe that these subsystems are
pairwise disjoint (supported in disjoint columns), and their
conjunction is equivalent to (\ref{1}).

We assume that the vector $v$ is provided in the following way (cf.
\cite{V}, \cite{R} and also above). A primitive real algebraic
element $u\in \overline{\QQ} \cap \RR$ is given as a root of a
polynomial $g\in \ZZ[Y]$ together with specifying a rational
interval $[e_1,e_2]$ which contains the unique root $u$ of $g$. In
addition, certain polynomials $g_j\in \QQ[Y],\, 1\le j\le n$ are
given such that $v_j=g_j(u)$. We suppose that $\deg(g), \deg(g_j)
\le d$ and that the absolute values of the numerators and
denominators of the (rational) coefficients of $g, g_j$ and of $e_1,
e_2$ do not exceed $2^M$.

To detect whether for a pair of the coordinates the congruence
$v_{j_1}-v_{j_2}\in \ZZ$ holds, the algorithm computes an integer
approximation $e\in \ZZ$ of $|v_{j_1}-v_{j_2}-e|< 1/2$ (provided
that it does exist) with the help of e.~g. the algorithm from
\cite{R} and then verifies whether $v_{j_1}-v_{j_2}=e$ exploiting
\cite{C}, \cite{G83} or \cite{R}. This supplies us with the
partition of the coordinates $v_1,\dots,v_n$ into the classes of
congruence. Thus, fixing for the time being the subsystem of
(\ref{1}) corresponding to a class of congruence, one can assume
w.l.o.g. that the coordinates of $v$ are integers. Moreover, one can
assume w.l.o.g. that $v=0$ replacing each $a_{i,j}$ by
$t^{-v_j}\cdot a_{i,j}$ (and keeping for them the same notations).

Then by elementary transformations with the rows of matrix $A$ over
the quotient-ring $\QQ(t)[Z]/h$ (making use of the basic algorithms
e.~g. from \cite{R}) and an appropriate permutation of columns, the
algorithm  brings $A$ to the form $a_{i,i}=1,\, a_{i,j}=0,\, 1\le
i\neq j\le m$ (one can assume w.l.o.g. that $rk (A)=m$).

For $m<j\le n$ denote $r_j:= - \min_{1\le i\le m} \{Trop
(a_{i,j})\}$. If $r_j < 0$ we put the coordinate $x_j=1$. Else if
$r_j \ge 0$ we put $x_j = y_{j,0}+y_{j,1}\cdot t + \cdots +
y_{j,r_j}\cdot t^{r_j}$ with the indeterminates
$y_{j,0},\dots,y_{j,r_j}$ over $\overline{\QQ}$.

For the time being fix $1\le i\le m$. Denote $s_i=\min _{m< j\le n}
\{Trop (a_{i,j}),\, Trop(b_i)\}$. The $i$-th equation of (\ref{1})
one can rewrite as \begin{eqnarray}\label{2} x_i+\sum_{m<j\le n}
a_{i,j}\cdot x_j =b_i
\end{eqnarray}
For every $s_i\le k\le 0$ one can express the coefficient of
$\sum_{m<j\le n} a_{i,j}\cdot x_j -b_i$ at the power $t^k$ as a
linear function $L_{i,k}$ over $\overline{\QQ}$ in the
indeterminates $Y:=\{y_{j,l},\, m<j\le n,\, 0\le l\le r_j\}$.

Consider the linear system \begin{eqnarray}\label{3} L_{i,k}=0,\,
1\le i\le m,\, s_i\le k< 0
\end{eqnarray}
in the indeterminates $Y$. The algorithm solves (\ref{3}) and tests
whether each of $n$ linear functions from the family
$$L:=\{L_{i,0},\, 1\le i\le m;\, y_{j,0},\, m<j\le n\}$$ \noindent does
not vanish identically on the space of solutions of (\ref{3}). If
all of them do not vanish identically then take any values of $Y$
which fulfil (\ref{3}) with non-zero values of all the linear
functions from the family $L$.
%and $L_{i,0}\neq 0,\,1\le i\le m$.
Then the equation (\ref{2}) determines uniquely $x_i$ with $Trop
(x_i)=0 (=v_i)$ and provides a solution $x$ of the system (\ref{1})
satisfying $Trop (x)=v$. Otherwise, if some of the linear functions
from the family $L$ vanishes identically on the space of solutions
of (\ref{3}) then the system (\ref{1}) has no solutions with $Trop
(x)=v$.

To test the above condition of identically non-vanishing of the
linear functions from the family $L$
%$L_{i,0},\, 1\le i\le m$,
the algorithm finds a basis $w_1,\dots,w_r \in (\overline{\QQ})^N$
and a vector $w\in (\overline{\QQ})^N$ where $N=|Y|$ such that the
$r$-dimensional space of solutions of (\ref{3}) is the linear hull
of the vectors $w_1,\dots,w_r$ shifted by the vector $w$. If each
linear function from the family $L$
%$L_{i,0},\, 1\le i\le m,\,y_{j,0},\, m<j\le n$
does not vanish identically on this space then all of them do not
vanish on at least one of the vectors from the family
$$F:=\{w+\sum_{1\le l\le R} p^l\cdot w_l,\, 1\le p\le nr+1\}$$ \noindent
because any linear function can vanish on at most of $r$ vectors
from $F$ due to the non-singularity of the Vandermond matrices.
%shifted by the vector $-w$ are linearly independent.
So, the algorithm substitutes each of  the vectors of $F$ into the
linear functions from  $L$
%$L_{i,0},\, 1\le i\le m$
and either finds a required one $Y$ or discovers that (\ref{1}) has
no solution with $Trop(x)=v$.

To estimate the complexity of the designed algorithm observe that it
solves the linear system (\ref{3}) of the size bounded by a
polynomial in $n,d$ with the coefficients from a finite extension of
$\QQ$ having the bit-size less than linear in $M$ and polynomial in
$n,d$ (again for the sake of simplifying the exposition a reader can
think just of the rational coefficients, cf. the remark above at the
beginning of the present Section). Thus, the algorithm solves this
system within the complexity polynomial in $M,n,d$ (see e.~g. \cite
{R}), by a similar magnitude one can bound the complexity of the
executed substitutions, and finally we can summarize the obtained
results in the following theorem.

\begin{theorem}\label{main}
There is an algorithm which for a tropical linear variety
$V:=\overline{Trop (P)}$ defined by a linear system (\ref{1}) over
the field $K$ of Puiseux series, recognizes whether a given real
algebraic vector $v\in ((\RR \cap \overline{\QQ}) \cup
\{\infty\})^n$ belongs to $V$. If yes then the algorithm yields a
solution $x\in K^n$ of (\ref{1}) with $Trop(x)=v$. The complexity of
the algorithm is polynomial in the bit-sizes of the system (\ref{1})
and of the vector $v$.
\end{theorem}

\section*{Further research}

Let a point $v\in \RR^n$ don't lie in a tropical linear variety
$V=\overline{Trop(P)}$ (cf. the Introduction). Theorem~\ref{main}
implies that one can verify the latter within the polynomial
complexity. Since \cite{SS}, \cite{S} entail that there exists a
finite tropical basis $f_1,\dots,f_k \in I$ of the ideal $I\subset
K[X_1,\dots,X_n]$ of $P$. This allows one to find $f\in
\{f_1,\dots,f_k\}\subset I$ such that $v$ is not a tropical zero of
the tropical polynomial $Trop (f)$. But the number $k$ can be
exponential (\cite{S}). Is it possible to construct $f_0\in I$ for
which $v$ is not a tropical zero of $Trop(f_0)$, within the
polynomial complexity?

What is the complexity to detect whether a given tropical linear
prevariety $V(R_1,\dots,R_p)$, with $R_1,\dots,R_p$ being tropical
linear polynomials, is a tropical variety?

\vspace{2mm} {\bf Acknowledgements}. The authors is grateful to the
Max-Planck Institut f\"ur Mathematik, Bonn for its hospitality
during writing this paper.

\end{document}